\documentclass[final,5p,times,twocolumn]{elsarticle}

\usepackage{enumitem}

\usepackage{subcaption}
\usepackage{dcolumn} 
\usepackage{booktabs}
\usepackage{multirow}
\usepackage{graphicx}

\usepackage{placeins} 

\usepackage[colorlinks=true,urlcolor=blue,breaklinks]{hyperref}

\usepackage{amsmath}
\usepackage{bm}
\usepackage[usenames, dvipsnames]{color}
\usepackage{ amssymb }

\usepackage[utf8]{inputenc}
\usepackage[english]{babel}

\graphicspath{{figs/}, {fig/}} 

\newcommand{\labr}{LaBr$_3$(Ce) }
\newcommand{\isotope}[2]{$^{#1}\mathrm{#2}$}
\newcommand{\pkg}[1]{\textsf{#1}}
\newcommand{\code}[1]{\texttt{#1}}

\usepackage{amssymb}

\usepackage{lineno}
\usepackage{wasysym}

\journal{NIM A}

\biboptions{sort&compress}

\biboptions{sort&compress}

\begin{document}
	
	\begin{frontmatter}
		
		
		
		\title{The $\gamma$-ray energy response of the Oslo Scintillator Array OSCAR}
		

		\author[]{F. Zeiser \corref{cor1}}
		\cortext[cor1]{fabio.zeiser@fys.uio.no}
		\author[]{G.M. Tveten \corref{add2}}
		\cortext[add2]{Present address: Expert Analytics AS, N0160 Oslo, Norway }
		\author[]{F.L. Bello Garrote}
		\author[]{M. Guttormsen}
		\author[]{A.C. Larsen}
		\author[]{V.W. Ingeberg}
		\author[]{A. Görgen}
		\author[]{S. Siem}

		\address{Department of Physics, University of Oslo, N-0316 Oslo, Norway}
		
		\begin{abstract}
			The new Oslo Scintillator Array (OSCAR) has been commissioned at the Oslo Cyclotron Laboratory (OCL). It consists of 30 large volume (\(\diameter\) 3.5 x 8 inches) LaBr$_3$(Ce) detectors that are used for $\gamma$-ray spectroscopy. The response functions for incident $\gamma$ rays up to 20 MeV are simulated with \pkg{Geant4}. In addition, the resolution, and the total and full-energy peak efficiencies are extracted. The results are in very good agreement with measurements from calibration sources and experimentally obtained mono-energetic in-beam  $\gamma$-ray spectra.
		\end{abstract}
		
		
	\end{frontmatter}
	
	\section{Introduction}
	The Oslo Cyclotron Laboratory (OCL) at the University of Oslo has commissioned the new Oslo SCintillator ARray (OSCAR) in 2018, replacing the NaI(Tl) scintillator array CACTUS \cite{Guttormsen1996}. The \labr detectors of OSCAR significantly improve the timing and energy resolution and intrinsic efficiency, which will not only provide better experimental conditions for the type of experiments most commonly carried out at OCL, but also open the door to novel studies. Since the early 1980's, the OCL has contributed substantially to experimental studies of nuclear properties in the quasi-continuum with the Oslo method \cite{Rekstad1983, Schiller2000, Larsen2011}. As the first step of the analysis it uses a common technique in nuclear physics and high energy physics called unfolding to calculate the emitted $\gamma$-ray spectrum from the measured spectrum \cite{Blobel2013, Choudalakis2012, Guttormsen1996}. This requires an accurate knowledge of the $\gamma$-ray response of OSCAR. In this article, we present simulations of the response and verify them by comparison to experimentally determined calibration sources and in-beam $\gamma$-ray spectra. The simulations are written in \pkg{C++} using the GEometry ANd Tracking 4 (\pkg{Geant4}) library \cite{Agvaanluvsan2004} v10.06, which is a standard tool for particle transport simulations in nuclear and  particle physics experiments. 
	The focus will be on the determination of the response between $\approx$100 keV and 10 MeV, as this is the energy range used in most common applications planed for OSCAR, the Oslo method and the study of prompt fission $\gamma$ rays.
	
	\section{Setup}
	OSCAR consists of a total of 30 large volume BriLanCe\texttrademark\, 380 LaBr$_3$(Ce) scintillating crystals manufactured by Saint-Gobain. The crystals are cylindrical with a diameter of 3.5 inches and a length of 8 inches. The detectors are coupled to Hamamatsu R10233-100 photo multiplier tubes (PMT) with active voltage dividers (LABRVD) \cite{Riboldi2011} and placed in an aluminum housing. They are powered by 6 \text{iseg NHS 60 20n} power supplies. The signals are processed by 14-bit, 500 MHz, XIA Pixie-16 modules and written to disk for subsequent offline analysis. An article dedicated to the detector characteristics and digital electronics will follow \cite{Ingeberg2020a}.
	
\begin{figure}[tbh]%
		\centering
		\begin{subfigure}[c]{0.9\columnwidth}
			\includegraphics[width=\textwidth]{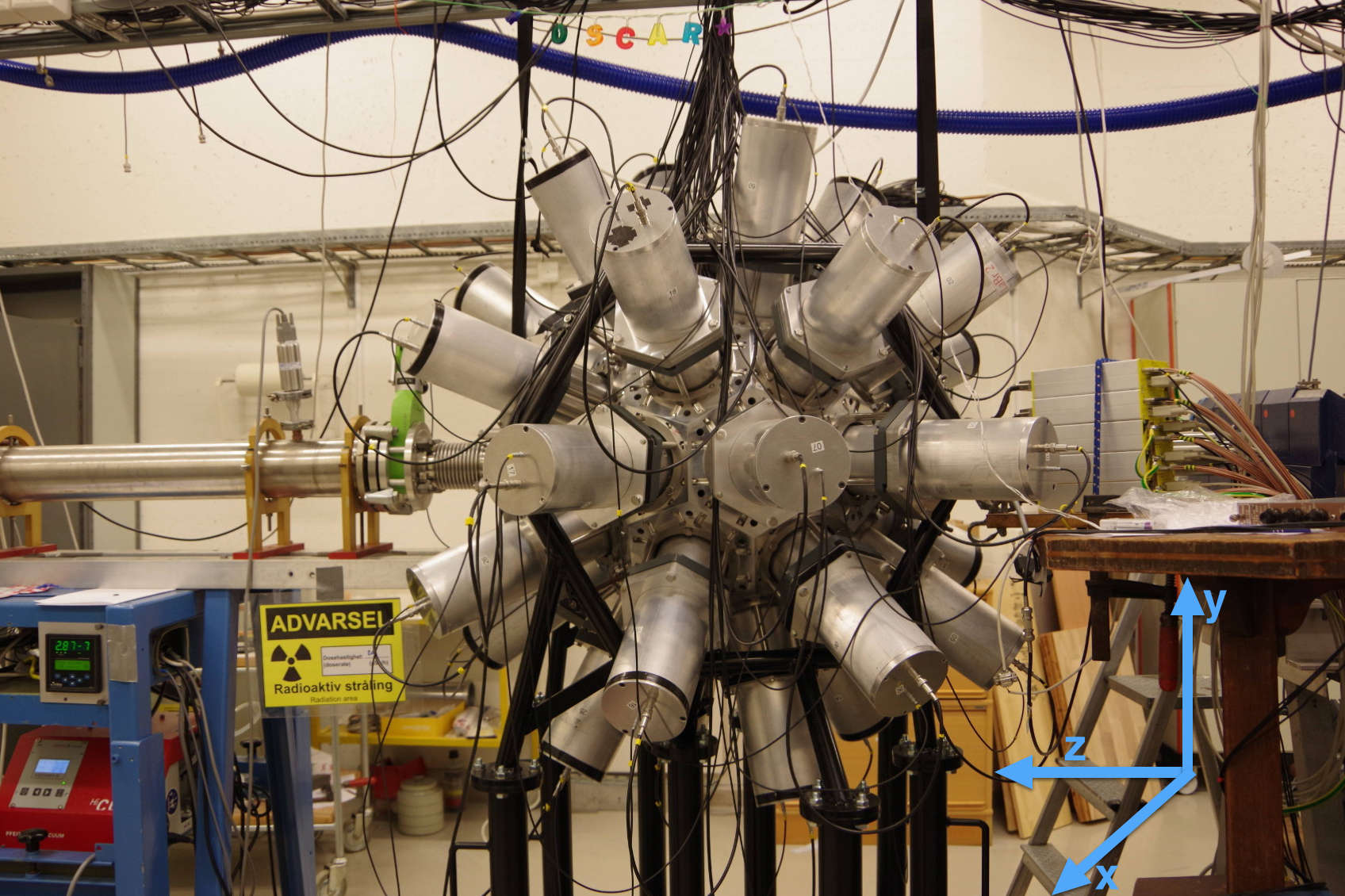}
			\caption{}
			\label{fig:OSCARdrawing}
		\end{subfigure}%
		\quad
		\begin{subfigure}[c]{0.9\linewidth}
			\includegraphics[width=\textwidth]{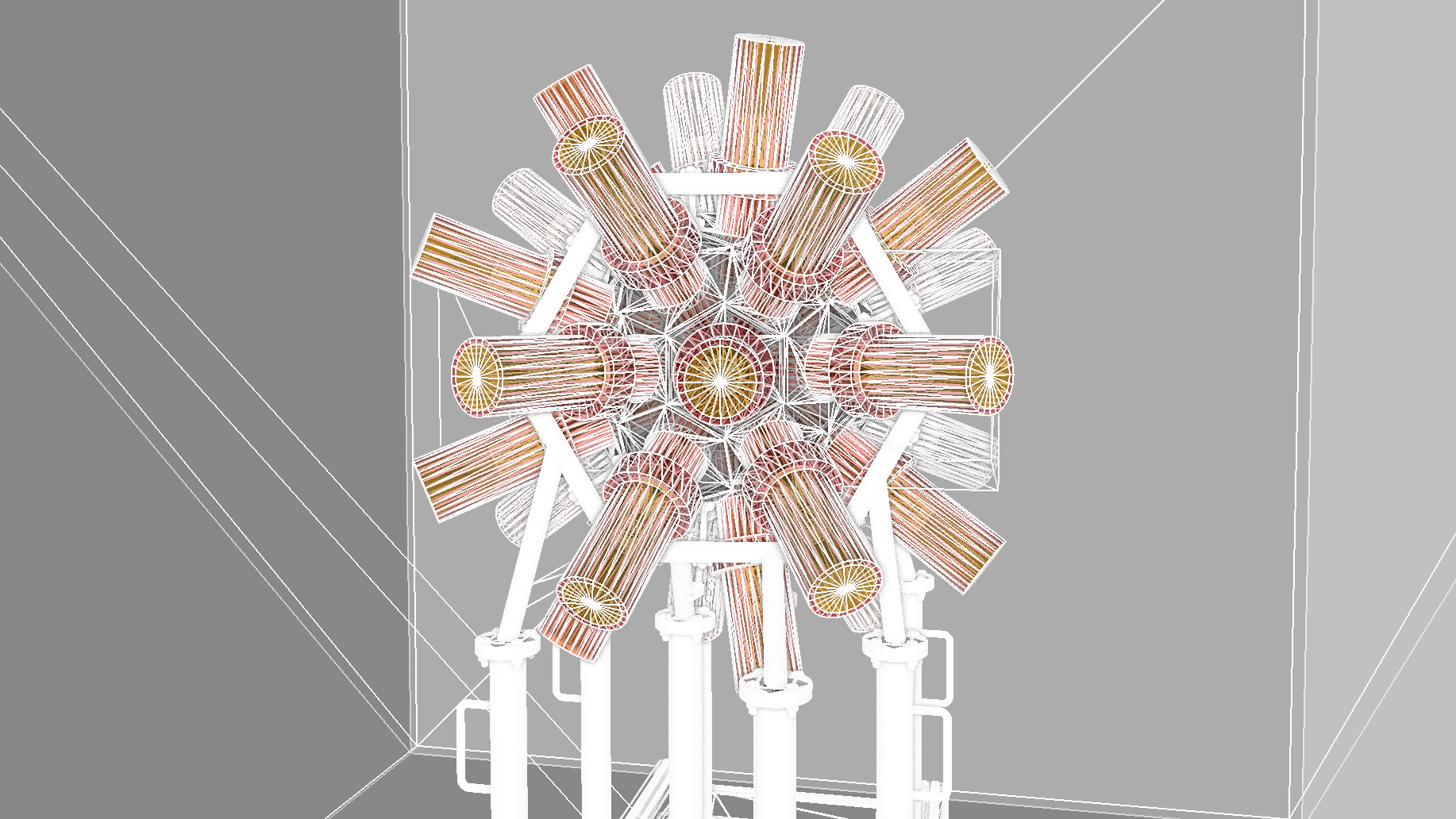}
			\caption{}
			\label{fig:OSCARGeant}%
		\end{subfigure}%
		\caption{a) A photo of the new array OSCAR and b) its geometry implementation in Geant4. The axes used in the Geant4 implementation are included as an overlay.}
		\label{fig:OSCARGeometry}%
	\end{figure}

	The detectors are mounted in a football shaped frame, see Fig.~\ref{fig:OSCARGeometry}, where the distance to the center is adjusted by the choice of distance rods. For this work, we used the closest configuration with a face-to-center distance $d$ between detector and source of $d \approx 16.3$ cm, which results in a solid angle coverage of 57\% of 4$\pi$.
	A table with the positions and labels of the OSCAR detectors is provided in \ref{app:angles}.
	
	The $\gamma$-ray detector array encompasses the beam-line and target chamber, in which the sources are placed, as well as the SiRi particle telescope \cite{Guttormsen2011} and for some experiments the NIFF PPAC fission fragment detector \cite{Tornyi2014}.
	
	\section{Implementation of the simulation}
	\subsection{Geometry}
	\label{sec:geometry}
	The implemented geometry includes the full setup of the array, including the OSCAR detectors and their support structure, SiRi, NIFF, two alternative target chambers, the beam-line, the target holder and the target frame or radioactive calibration source itself. The full model v2.0.0 is available on \href{https://github.com/oslocyclotronlab/OCL_GEANT4}{github} and as Ref.\,\cite{Zeiser2020a}. Most components can be (de)activated via macros at runtime to reflect the experimental conditions or to speed up the calculations. The standard configuration of the experiments is available via \path{setup_normal_run.mac} and does not include NIFF and the calibration source, as in-beam spectra on very thin metal foils are used. By default, we use the spherical target chamber installed in 2018. 
	
	To maintain a high performance of the simulations we have used the Constructed Solid Geometry (CSG) wherever feasible. Thus, the radioactive source, the detectors including their encapsulation and housing, and the football-like shaped aluminum frame, a truncated icosahedron, are implemented as CSG solids. The polar angle $\theta$ and azimuth angle $\phi$ of the 30 detectors are fixed by the truncated icosahedron geometry and given in Table \ref{tab:angles} in the Appendix; the beamline runs through the remaining two faces. As common practice, the z-axis is chosen along the beam direction, and the y-axis points upwards. The face-to-center distance $d$ between detector and source can be physically adjusted by different spacer rods; in the simulations $d$ can be adjusted for each detector individually with a macro command, or they may even be removed totally, which facilitates updating the response matrix for experiments in a non-standard configuration. By default, all 30 detectors are placed at a distance of $d\approx16.3$ cm.
	
	An older cylindrical target chamber is  dedicated to actinide experiments and is implemented via CSG solids. The new spherical target chamber, including the wheel with the target holders, has a much more complex geometry, such that we used the Computer-Aided Design (CAD) drawings instead. Similarly, the support structure of the frame is implemented with the CAD geometry.
	
	All CAD drawings are imported as \pkg{GDML} files after conversion with \pkg{GUIMesh} v1 \cite{Pinto2019}. We preprocessed the drawings slightly, removing several small elements like sealing rings, which are not expected to effect the $\gamma$ rays significantly, but may increase the computation time considerably. Each element of the setup is implemented through an individual \pkg{Geant4} \textit{parallel world geometry} to facilitate the navigation and avoid boundary problems. The \textit{layered mass geometry} ensures that a particle at any given point only sees the topmost parallel world with a volume defined at the point, or if no parallel world is defined it seems the basic \textit{mass world}. We use following top to bottom hierarchy: 
	\begin{enumerate}[nosep]
		\item \code{ParallelWorld Targets on Wheel}: Target frames placed on the target wheel,
		\item \code{ParallelWorld SiRi}: A CAD implementation of SiRi particle telescopes (a more primitive CSG implementation exists),
		\item \code{ParallelWorld Frame Outer}: The support structure of the frame,
		\item \code{ParallelWorld Target Chamber}: The spherical target chamber including the target wheel,
		\item \code{massWorld} The \textit{normal} world, where all CSG volumes are defined.
	\end{enumerate}
	
	The rectangular calibration sources are modeled given the manufacturer specifications of Eckert \& Ziegler through a 0.5 $\mathrm{cm}^3$ acrylic glass cube of the support material, embedding an Amberlite\texttrademark\, IR-120 sphere containing the active material with a radius of 0.5 $\mathrm{mm}$ . 
	
	All commands related to the geometry are in the \path{/OCL} macro directory and its sub-directories and are documented online. The geometry is constructed in a modular fashion, such that it is easy to reuse parts of the code when the \labr detectors are used at another experimental facility.
	
	\subsection{Physics processes and event generation}
	We have chosen the \pkg{QGSP\_BIC\_HP} reference physics list, which implements standard electromagnetic interactions through \pkg{G4EmStandard\-Physics} and neutron interactions through data driven high precision cross-sections. All events are generated at the target position at the center of the sphere using the \pkg{General Particle Source} (GPS). For the calibration sources we use the \pkg{Radioactive Decay Module} in addition. Whilst the active area of the calibration source is approximated by a small spherical source of the carrier material, we used a small ($\approx 3 \mathrm{mm}^2$) isotropic planar source roughly corresponding to the beam size for the cyclotron at the target position. The scintillation process can be simulated with \pkg{G4OpticalPhysics}, but is not included in the simulations by default, as it significantly increases the computation time without any impact on the energy collection.

	\begin{figure*}[tb]
		\begin{center}
			\includegraphics[clip, trim=0 0 35 25, width=\columnwidth]{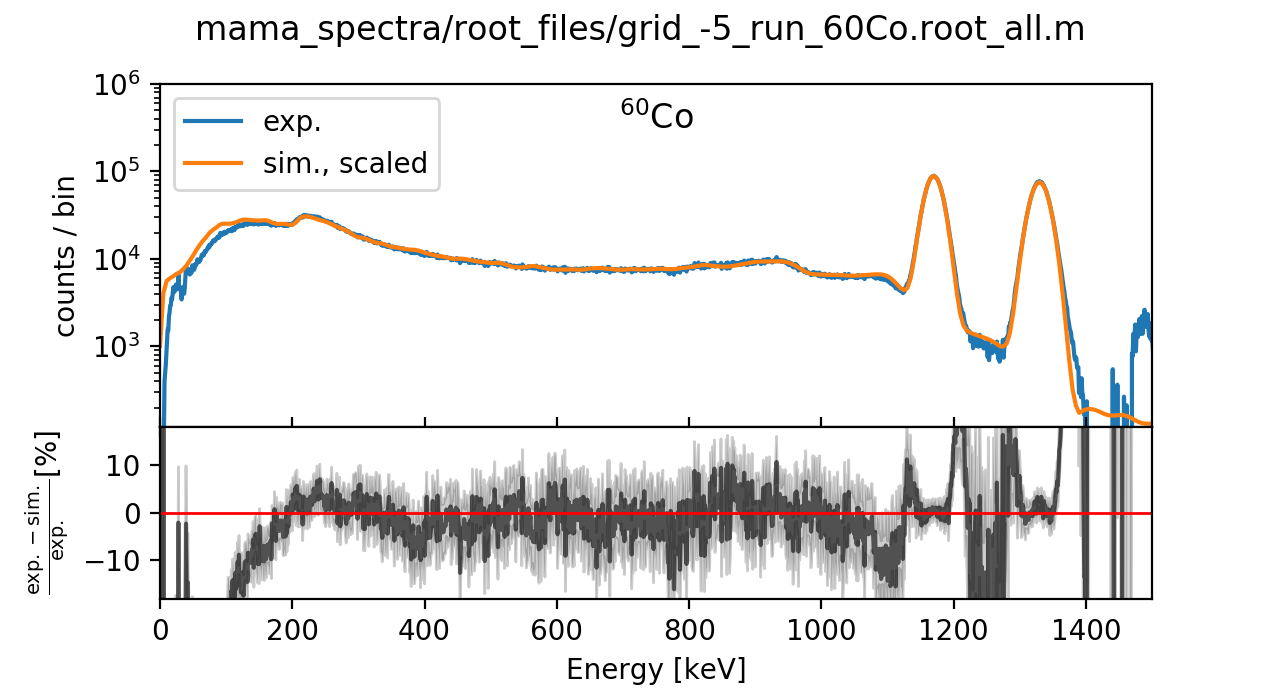}
			\includegraphics[clip, trim=0 0 35 25, width=\columnwidth]{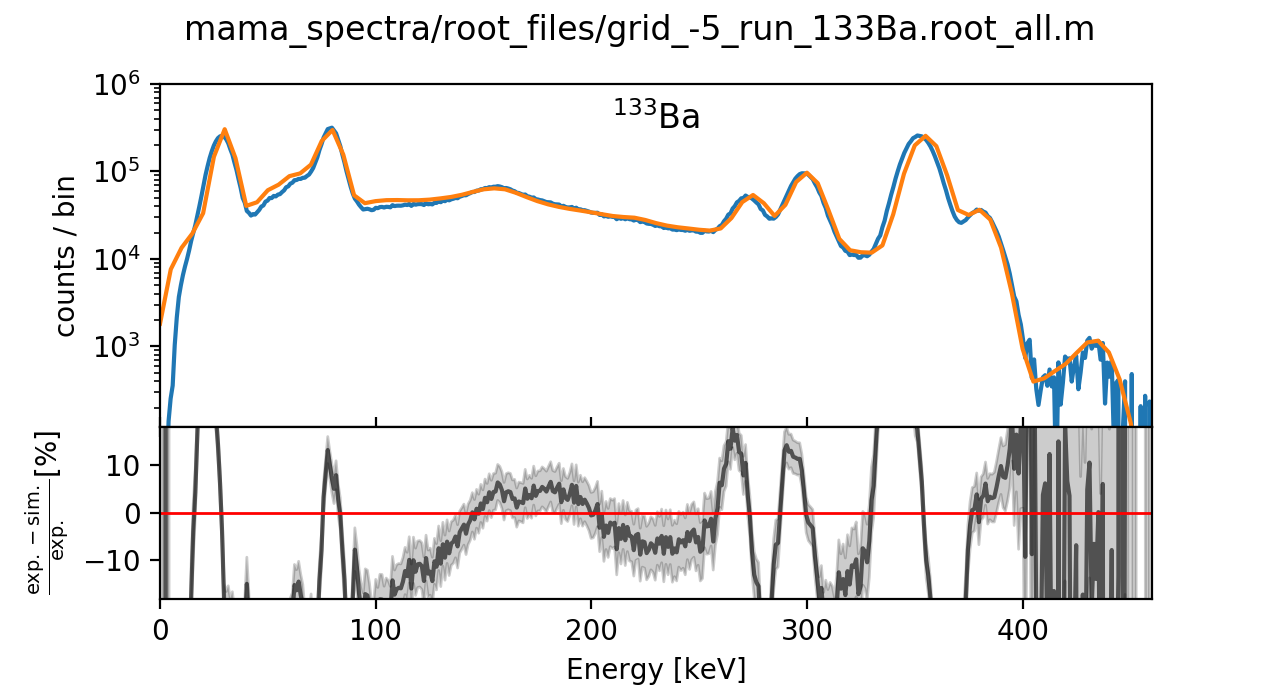}
			\includegraphics[clip,trim=0 0 35 25,width=\columnwidth]{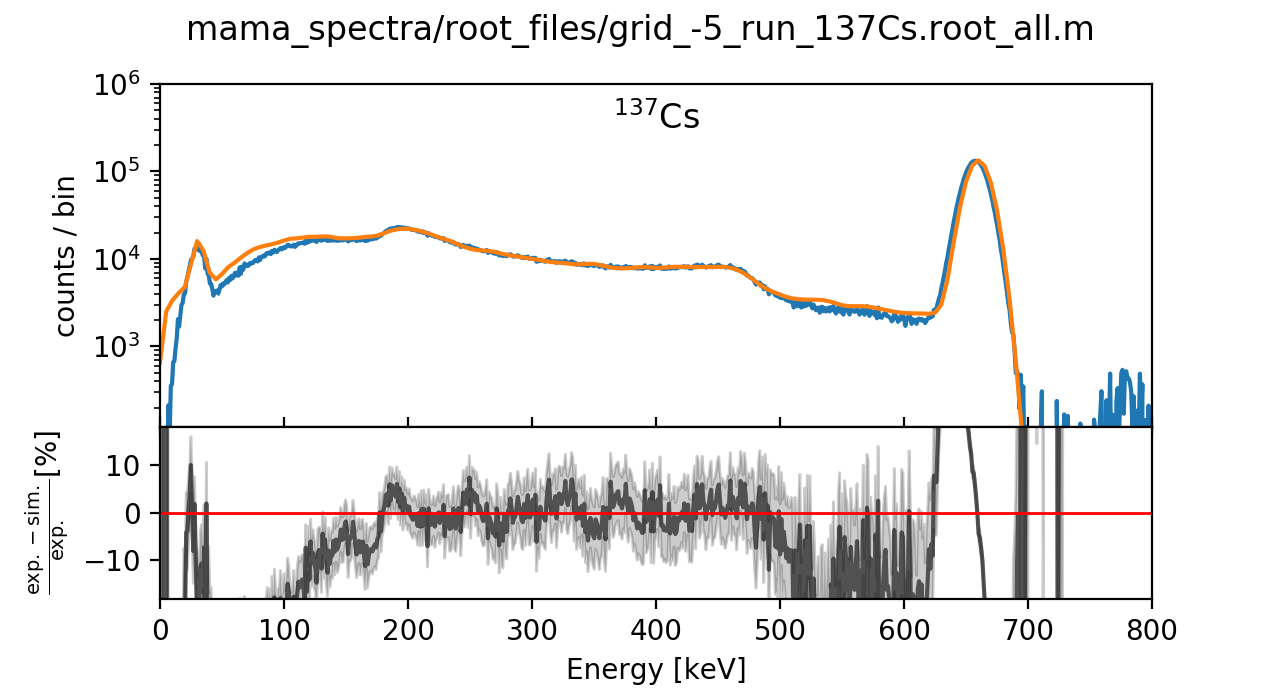}
			\includegraphics[clip,trim=0 0 35 25,width=\columnwidth]{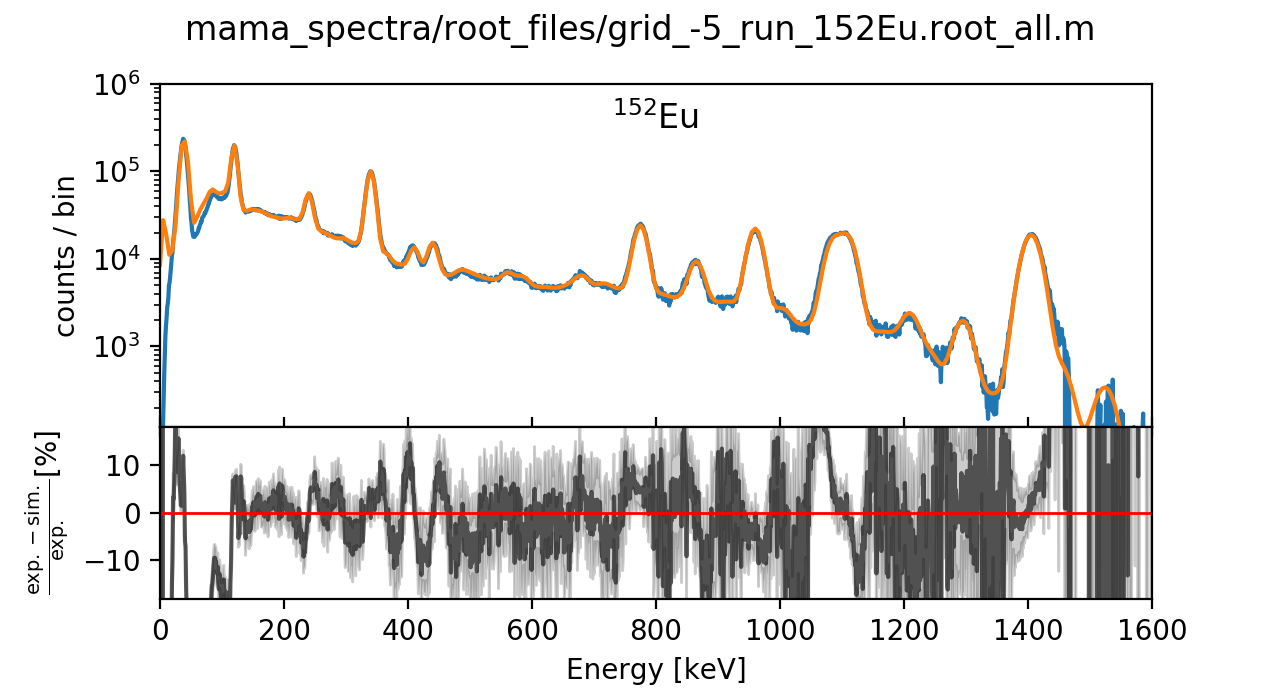}
					
			\caption{Experimental source spectra (blue) compared to the simulations (orange) for \isotope{60}{Co} (top left), \isotope{133}{Ba} (top right), \isotope{137}{Cs} (bottom left) and \isotope{152}{Eu} (bottom right). The simulations are scaled to the same number of decays as the experimental spectra. The lower panels show the relative difference between experiment. Note that the displayed energy ranges are different for each isotope.}
			\label{fig:sources}
		\end{center}
	\end{figure*}
	
	\subsection{Scoring and data analysis}
	\label{sec:scoring-and-data-analysis}
	The energy deposited in each crystal is recorded as an n-tuple and stored as a \pkg{ROOT} \cite{Brun1997} tree. For the further analysis, we combine the histograms for all detectors to a cumulative response of OSCAR. As \pkg{Geant4} does not model the electronic response of the system and inclusion of the scintillation process for large scale simulations is prohibited by the computation time, we initially get histograms with spikes at the full-energy peak, single escape, etc.. These are folded by a Gaussian to mimic the statistical behavior and non-uniformity of the scintillation photon collection, the PMT and the signal processing electronics. The full width at half maximum (FWHM) is determined by fits to 15 peaks from following radioactive sources: \isotope{60}{Co}, \isotope{133}{Ba}, \isotope{137}{Cs}, \isotope{152}{Eu} and \isotope{241}{Am}. The variation of the FWHM as a function of the $\gamma$-ray energy $E_\gamma$ is fitted by
	\begin{align}
	\mathrm{FWHM}(E_\gamma)\,[\mathrm{keV}]= \sqrt{a_0 + a_1 E_\gamma + a_2 E_\gamma^2},
	\end{align} 
	with the best fit values $a_0 = 60.64(73)$, $a_1 = 0.458(02)$, and $a_2 = 2.6555(17)\times10^{-4}$, assuming $E_\gamma$ is given in keV.
	
	\section{Results and verification}
	\subsection{Calibration sources}
	\label{sec:source}
	The response has been simulated for $2\times10^5$ decays of \isotope{60}{Co}, \isotope{133}{Ba}, \isotope{137}{Cs} and \isotope{152}{Eu} (the activity of the \isotope{241}{Am} is not known) and is compared to the experimental measured spectra in Fig.\,\ref{fig:sources}. The simulations are scaled to same number of decays as in the experimental data, using the activity and measuring time of the sources. We use a quadratic energy calibration fitted to 15 peaks of all four sources and subtract the background. This generally works quite well, although small deficiencies are visible in the  \isotope{60}{Co} and \isotope{152}{Eu} spectra between $\approx$1.4 and 1.5 MeV, where the strong internal radiation of the \labr detectors is not correctly subtracted due to a small drift of the calibration. The simulations give an excellent reproduction of the calibration spectra with an average deviation below 5\% for $\gamma$-ray energies $E_\gamma$ above 200 keV. 
	
	Between 50 and 200 keV the deviations reach up to about 20\%. We expect larger deviations in this area, as low energy $\gamma$ rays are easily attenuated, thus require an even more precise implementation of the geometry. In the most common application of OSCAR, the Oslo method, usually $\gamma$ rays between $\approx$1 and 10 MeV are studied. Newer applications, like the study of prompt fission $\gamma$ rays, may also study the $\gamma$-ray response above $\approx$100 keV. However, when the low energy region ($\leq 100$ keV) is of interest, it might be important to include details like the cables leading to SiRi and the target wheel, small screws, as well as sealing rings, which were removed from the CAD drawings to improve the computation time, see Sec.\,\ref{sec:geometry}. Furthermore, non-linearities in the detector response will become relevant at these energies. It can also be noted that the magnitude of the deviation depends on the exact source spectrum, thus we cannot find a generally applicable correction function that could be used on top of the simulations. Below $\approx$30 to 50 keV, the above considerations on the geometry are valid as well, but in addition the onset of the detector threshold (which cannot be simulated with \pkg{Geant4}) will lead to a loss of counts in the experimental spectra.
	
\begin{figure}[tb]
		\begin{center}
			\includegraphics[clip,trim=0 0 45 30,width=\columnwidth]{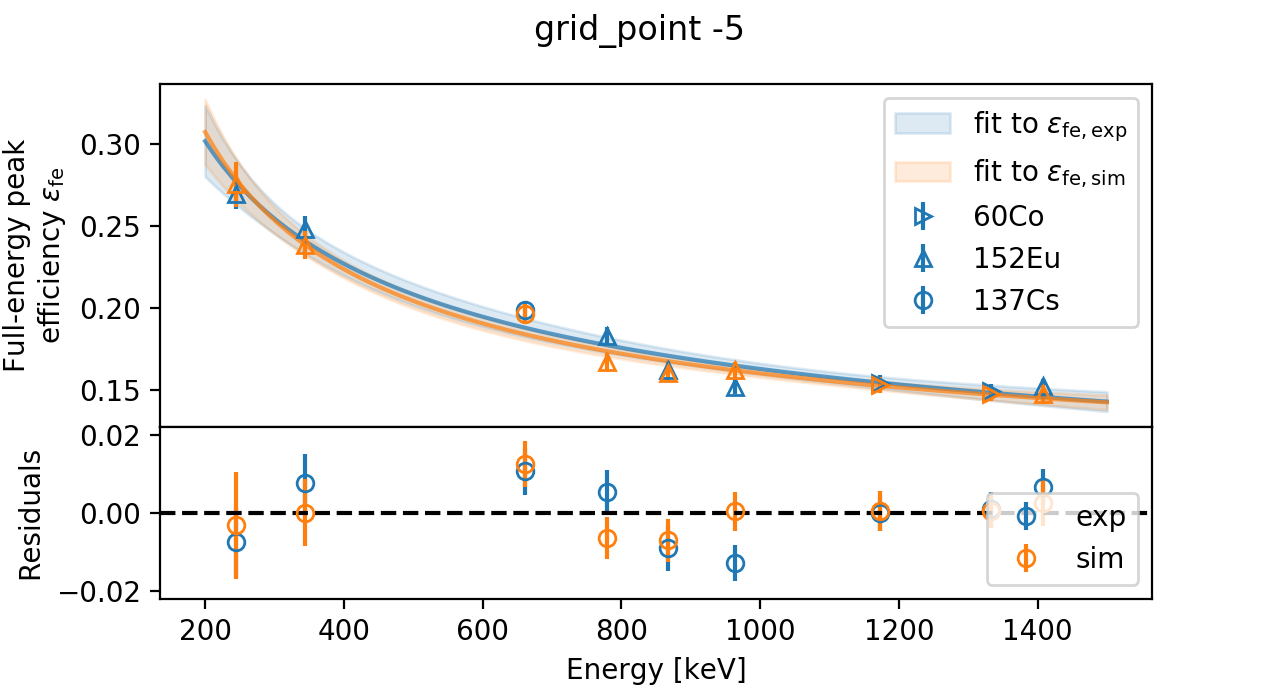}
			\caption{Full-energy peak efficiency of OSCAR from calibration sources (blue) and simulation (orange). The bottom panel shows the residuals from the fits.}
			\label{fig:photoeff}
		\end{center}
	\end{figure}

	The full-energy peak efficiency $\epsilon_\mathrm{fe}$ of the setup has been analyzed and is displayed in Fig.\,\ref{fig:photoeff}. The results depend on the choice of the fit function and minimization routine, and we estimate this systematic uncertainty to $\approx 3-5\%$. Our main goal here is to verify the simulated efficiencies. Therefore we fit the peaks in both the experimental and simulated spectra in the exact same way, using a fit function composed of following three elements:
	\begin{enumerate}[nolistsep]
		\item a Gaussian: The main component of the peak arising from the electronic response to an otherwise mono-energetic $\gamma$ ray,
		\item a smoothed step function: From Compton scattered $\gamma$ rays that enter the detector and the escape of scintillator photons from the crystal, using the functional form proposed in \pkg{RadWare}'s \pkg{gf3} \cite{gf3},
		\item an constant background: From the contribution of other peaks and their Compton spectrum.
	\end{enumerate} 
	We attempted to extend the fitting routine for more complicated peak structures, like the double peaks in \isotope{133}{Ba} and \isotope{152}{Eu}, but could not find a good and consistent way of fitting them. For the comparison in Fig.\,\ref{fig:photoeff}, only fits that converged properly are processed. The full-energy peak efficiencies $\epsilon_\mathrm{fe}$ agree within the fit uncertainties. 
	The analyzed efficiencies $\epsilon_\mathrm{fe}$ in turn are fit as proposed in \pkg{gf3} \cite{gf3}:
	\begin{align}
	\ln \epsilon_\mathrm{ph} = p_0 + p_1 \ln(E_\gamma) + p_2 \ln(E_\gamma)^2.
	\label{eq:photoeff}
	\end{align}
	The best-fit values are listed in Tab. \ref{tab:fiteff}.
	
	\subsection{In-beam spectra}
	\label{sec:in-beam}
	The comparison to experimental in-beam spectra provides additional insights, but also challenges to the data processing. 
	
	\begin{figure}[tbh]%
		\centering
		\begin{subfigure}[c]{\columnwidth}
			\includegraphics[clip,trim=0 0 32 25,width=\textwidth]{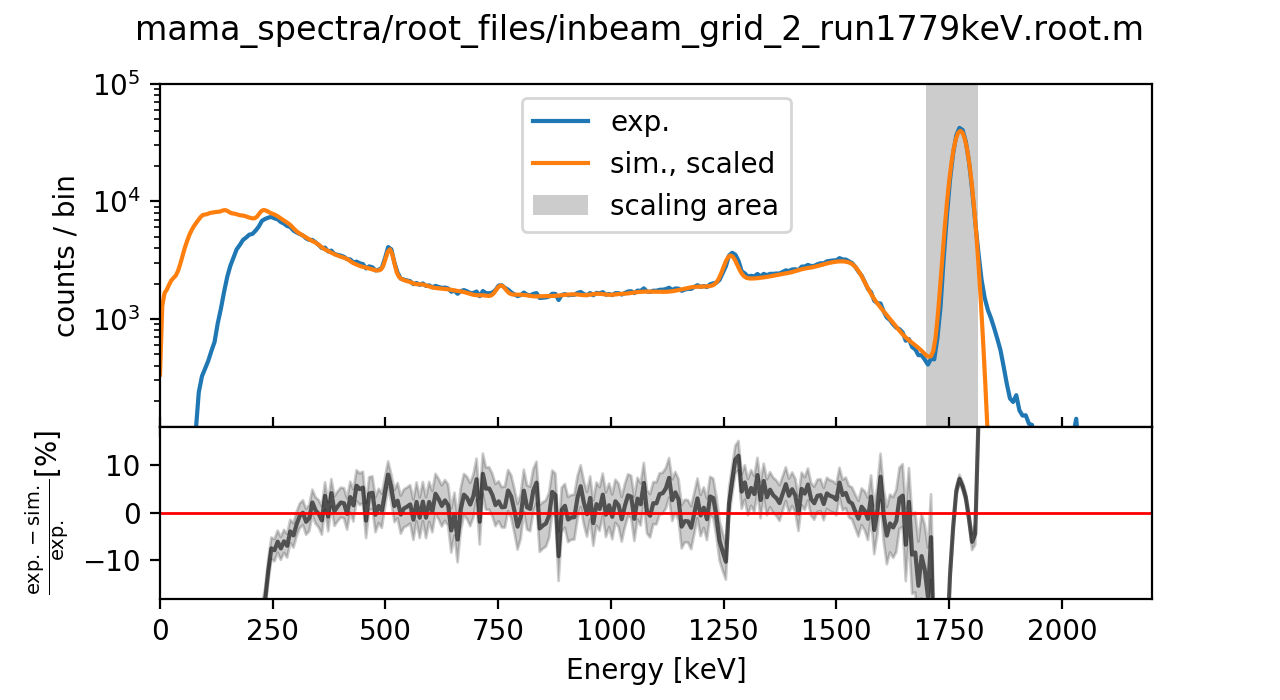}
		\end{subfigure}%
		\quad
		\begin{subfigure}[c]{\columnwidth}
			\includegraphics[clip,trim=0 0 32 25,width=\textwidth]{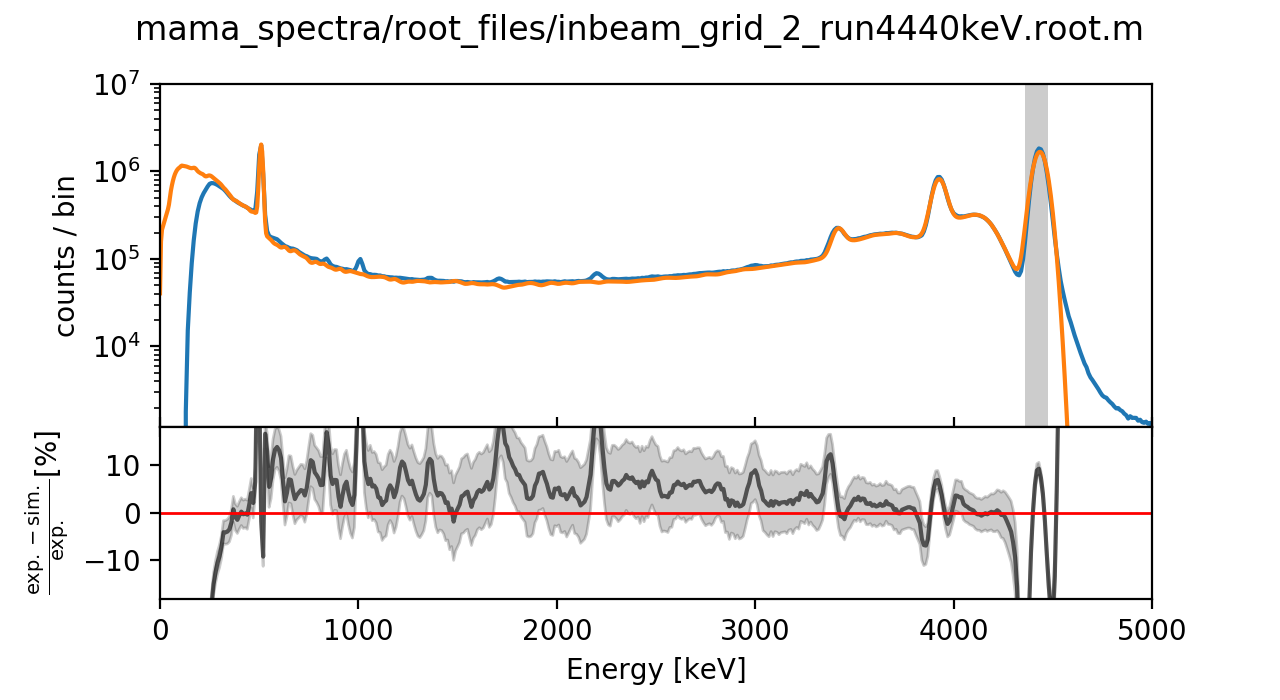}
		\end{subfigure}%
		\caption{Experimental in-beam $\gamma$-ray spectra (blue) compared to simulations (orange) of mono-energetic incident $\gamma$ ray of 1779 (top) and 4440 keV (bottom), respectively. The lower panels show the relative difference between experiment and simulation. The area used to scale the simulations to the experiment is highlighted in gray. The discrepancies are explained in the text.}
		\label{fig:inbeam}%
	\end{figure}
	
	We use particle-$\gamma$ coincidence measurements from the (p,p')\isotope{28}{Si} and (p,p')\isotope{12}{C} reactions, both measured in 2019. Gating on the detected particle energy, we can select $\gamma$ rays from population of specific excited states and their decay, e.g.\,the first excited states at 1779 and 4440 keV, respectively. It was not possible to obtain other mono-energetic spectra from \isotope{28}{Si} by gaiting on higher excitation, because the particle energy resolution was not good enough to clearly distinguish the different states (and separate them from states of \isotope{29}{Si} and \isotope{16}{O} contaminating the target).
	
	For the in-beam spectra random coincidences with particles from previous beam bursts of the cyclotron are subtracted using two dimensional graphical time-energy cuts. The total number of incident $\gamma$ rays is not known for the experimental spectra. Therefore, we use the number of detected $\gamma$ rays that fulfill the coincidence requirements and normalize the simulated and experimental spectra to the same number of counts in the full-energy peak. Naturally, the spectra agree within the close vicinity of the full-energy peak, but from Fig.\,\ref{fig:inbeam} we can see that they also match well for the single- and double-escape peaks, the annihilation peak and the Compton-spectrum above $E_\gamma \approx 200$ keV. 
	
	There are several noteworthy discrepancies. Below $E_\gamma \approx 200$, the simulations seemingly overestimate the experimental spectra significantly. We attribute this to an over-subtraction of the background for low energies in connection with inaccurate graphical time-energy gates. On the contrary, the simulations apparently underestimate the Compton-background for the 4440 keV $\gamma$-ray spectrum from \isotope{12}{C}. This is however linked to the  contamination of $\gamma$ rays from the aluminum frame of the \isotope{12}{C} target, which contaminate the experimental \isotope{12}{C} spectrum. This can be identified from the position of the additional aluminum peaks, where the 1014, 1720, and 2212 keV lines are easily visible in Fig.\,\ref{fig:inbeam}. Finally, we observe that the full-energy peak in the experimental spectra is not perfectly Gaussian shaped, but has a tail towards higher energies. Note that this is not visible in the source spectra of Fig.\,\ref{fig:sources}. It is beyond the scope of this article to verify whether the tail is due to suboptimal settings during the data acquisition (e.g.\,different impedances causing a slight ringing in the cables) such that it can be removed in future experiments, or whether it is of permanent nature (e.g.\,pileup with $\gamma$ rays or x-rays created from the cyclotron operation, etc.). In the latter case, one could use a non-Gaussian kernel for the smoothing of the simulated spectra, which is described in Sec.\,\ref{sec:scoring-and-data-analysis}.
	
	\subsection{Response matrix}
	For the previously used CACTUS $\gamma$-ray detectors, the response matrix was obtained from an interpolation and extrapolation of a small number of measured experimental spectra \cite{Guttormsen1996}. Given the simulations presented here and their successful verification in Sec.\,\ref{sec:source} and \ref{sec:in-beam}, we now calculate the response of OSCAR for a large grid of incident $\gamma$-ray energies $E_\mathrm{in}$ between 50 keV and 20 MeV. A simulation of $10^6$ single incident $\gamma$-rays of 5 MeV in the standard setup with the new spherical target chamber takes about 8 cpu-hours on a single Intel E5-2683v4 2.1 GHz core. As most experiments only require the response below 10 MeV, we split the calculations in two parts. Below 10 MeV we use a step-size of 10 keV, removing the need for interpolations; in addition we simulate the response for 12, 15 and 20 MeV. As an additional measure to balance runtime against the accuracy of the results, we increase the number of events from $0.5\times10^6$ for the low energies to $3\times10^6$ for the highest incident energies. The total computation time of the response was $\approx 17000$ cpu-hours and it is available online on \href{https://github.com/oslocyclotronlab/OCL_GEANT4}{github} and as dataset Ref.\,\cite{Zeiser2020a} in the matrix format $R(E_\mathrm{in}, E_\mathrm{out})$, where $E_\mathrm{out}$ is the simulated response. Note that we use an isotropic source with multiplicity 1 for all events here, but the source definition can easily be adopted for higher multiplicities and other angular distributions through the GPS macro commands if this is desirable.
	
\begin{figure}[tbh]%
		\centering
		\includegraphics[width=0.9\columnwidth]{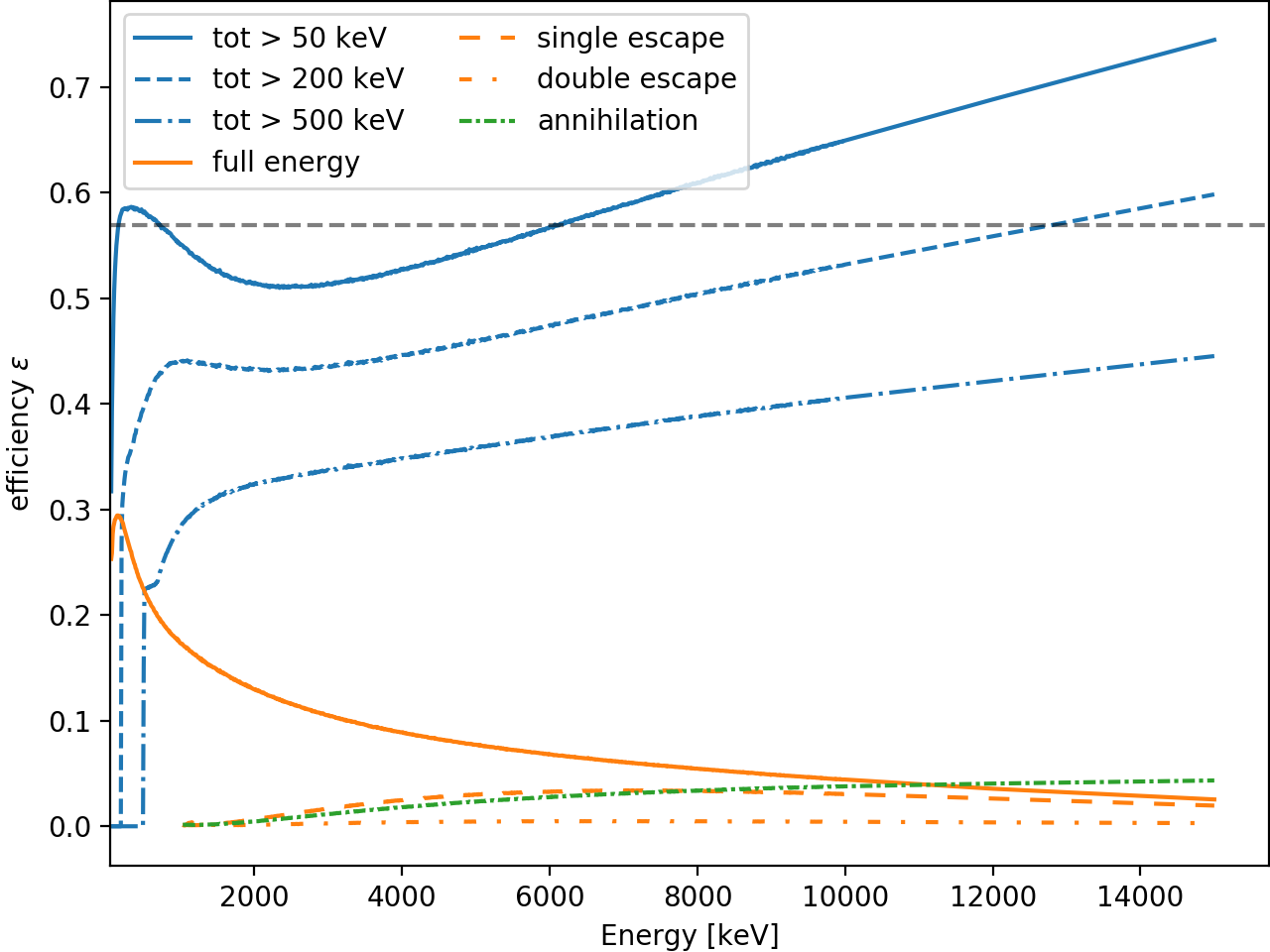}
		\caption{Simulated efficiencies (see legend), where the total efficiency is given for different lower thresholds. The geometric efficiency of $57\%$ is highlighted (black dashed line), but can be exceeded due e.g. cross-talk between detectors for one and the same gamma event.}
		\label{fig:efftot}
	\end{figure}

	In Fig.\,\ref{fig:efftot} the total efficiency is plotted, which is given by the ratio of counts detected above a threshold over the number of simulated events. As the most common unfolding technique that is used in the Oslo method \cite{Guttormsen1996} requires the full energy, single and double escape, and annihilation probabilities for the so called Compton subtraction method, these probabilities are extracted as well.
	
	\section{Lessons learnt}
	A first version of the OSCAR simulation was developed in 2018 \cite{Zeiser2018}, but we encountered several challenges in the model development and the comparison to experiments \cite{Zeiser2020}. In the following, we try to summarize the main lessons learnt which lead to the very good agreement between simulation and experiment.
	
	\begin{enumerate}
		\item As mentioned in Sec.\,\ref{sec:source}, the full-energy peak efficiency is rather sensitive to the fit function and procedure. In the simulations, it is possible to select only photoeffect interactions and base the full-energy peak efficiency on these. However, this induced a systematic discrepancy and lead to an apparently poorer reproduction of the experimental fits.
		
		\item Several studies have shown that the \labr detectors have a non-linear energy response, especially at low energies, see e.g.\,Refs.\,\cite{Gosta2018} and references therein. However, during the first benchmarking phase for the simulations, only a subset of the calibration sources was used, which only allowed for a linear energy calibration. This induced an error which was misattributed to the accuracy of the geometry implementation. Ideally, even more calibration sources should be available if one wants to improve the response below $E_\gamma = 200$ keV.
		
		\item Initially, we experienced large problems importing the CAD geometry, with particle tracks getting stuck. This was easily resolved with \pkg{GUIMesh} and the parallel world geometry. Problems with the material definition in \pkg{GUIMesh} v1 were circumvented by grouping the elements of a drawing by material, exporting each group individually, and editing the material through a \textit{search and replace}. 
	\end{enumerate}
	
	\section{Summary}
	
	Response functions of the new $\gamma$-ray detector array OSCAR at the OCL have been simulated with the \pkg{Geant4} toolkit up to 20 MeV. The simulations are compared to experimental spectra from calibration sources and in-beam $\gamma$-rays, where a good agreement has been achieved. The deviations are below $\approx 5\%$ for $\gamma$-ray energies $E_\gamma$ larger than 200 keV. Additionally, we obtained the total and partial efficiencies for the various components of the $\gamma$-ray interaction with the detectors. Finally, we summarized several of the main challenges of the analysis.
	
	\section*{Acknowledgments}
	The scientific advisory board of the OSCAR project, consisting of Franco Camera, David Jenkins and Pete Jones, have been important for the OSCAR project and we wish to thank them for their valuable advice. We would also like to thank Jan Mierzejewski and C3D for the mechanical design of the frame and Agnese Giaz for spending time transferring knowledge on LaBr$_3$(Ce) scintillator detectors to us at OCL. The fact that the INFN Milan lended us a Labrpro unit was a great help to the project. 
	We would like to thank Kevin C.W. Li for his excellent comments on the \pkg{Geant4} simulations, and Wanja Paulsen for providing the calibrated experimental \isotope{12}{C} spectrum. We wish to thank the OCL engineers, Jan C. Müller, Jon Wikne, Pawel A. Sobas and Victor Modamio for excellent experimental conditions for testing OSCAR. The OSCAR Array is financed by the Research Council of Norway under contract no.\,245882. F.~Z., V.~W.~I., A.~G. and S.~S. acknowledge funding from the Research Council of Norway under contract no. 263030, G.~M.~T acknowledges funding under contract no.\, 262952. 
	A.~C.~L. acknowledges funding from the European Research Council through ERC-STG-2014 under grant agreement no.\,637686, and support from the “ChETEC” COST Action (CA16117), supported by COST (European Cooperation in Science and Technology). The grid computations were performed on resources provided by UNINETT Sigma2 the National Infrastructure for High-Performance Computing and Data Storage in Norway.
	
	\section*{CRediT authorship contribution statement}
	Conceptualization, Software, Methodology: F.Z. and G.M.T.;	
	Validation, Writing - Original Draft, Visualization: F.Z.;	
	Formal analysis and Investigation: F.Z. and F.L.B.G.;	
	Data Curation : F.Z., M.G. and V.W.I.;	
	Writing - Review \& Editing: F.Z., G.M.T, F.L.B.G., M.G., A.C.L.;	
	Funding acquisition: A.G., A.C.L., and S.S.

	\vfill\eject  
	\appendix
	\section{Tables}

	\label{app:angles}
	\begin{table}[htb]
		\caption{Labeling of the detectors and the position of the detectors determined by the geometry of the frame. This labeling is used both in the Geant4 simulations and on the actual detector frame of OSCAR. The angles $\theta$ and $\phi$ specify the detector location in spherical coordinates, the distance can be varied by spacers.}
		\begin{tabular}{ l c c c }
			\hline
			Ring & Det. number  & $\theta$ [deg]  & $\phi$ [deg]  \\
			\hline
			1  & 1      & 142.6       & 0.0     \\
			& 2      & 142.6       & 72.0      \\
			& 3      & 142.6       & 144.0     \\
			& 4      & 142.6       & 216.0     \\
			& 5      & 142.6       & 288.0     \\ 
			2  & 6      & 116.6       & 324.0     \\ 
			& 8      & 116.6       & 36.0      \\ 
			& 10     & 116.6       & 108.0     \\ 
			& 12     & 116.6       & 180.0     \\ 
			& 14     & 116.6       & 252.0     \\ 
			3  & 7      & 100.8       & 0.0     \\ 
			& 9      & 100.8       & 72.0      \\ 
			& 11     & 100.8       & 144.0     \\    
			& 13     & 100.8       & 216.0     \\ 
			& 15     & 100.8       & 288.0     \\ 
			4  & 16     & 79.2        & 324.0     \\
			& 18     & 79.2        & 36.0      \\
			& 20     & 79.2        & 108.0     \\
			& 22     & 79.2        & 180.0     \\
			& 24     & 79.2        & 252.0     \\
			5  & 17     & 63.4        & 0.0     \\
			& 19     & 63.4        & 72.0      \\
			& 21     & 63.4        & 144.0     \\
			& 23     & 63.4        & 216.0     \\
			& 25     & 63.4        & 288.0     \\
			6    & 26     & 37.4        & 324.0     \\
			& 27     & 37.4        & 36.0      \\
			& 28     & 37.4        & 108.0     \\
			& 29     & 37.4        & 180.0     \\
			& 30     & 37.4        & 252.0     \\
			\hline
		\end{tabular}
		\label{tab:angles}
	\end{table}
	
	\begin{table}[tb]
		\centering
		\caption{full-energy peak efficiency fit coefficients, see Eq.\,\ref{eq:photoeff} and their covariances. Note that the values are strongly correlated, thus the efficiency is better determined as it might seem from the quoted $1\sigma$ uncertainties for each parameter.}
		\label{tab:fiteff}
		
		\begin{subtable}{\columnwidth}
			\centering
			\caption{Optimal parameters}
			\begin{tabular}{r|rrr} \toprule
				& \multicolumn{1}{c}{$p_0$} & \multicolumn{1}{c}{$p_1$} & \multicolumn{1}{c}{$p_2$} \\ \midrule
				exp. & 1.94(27) & 0.75(85) & 0.030(66) \\ 
				sim. & 3.0(21) & -1.11(65) & 0.058(50) \\ \bottomrule
			\end{tabular} 
			\label{tab:photo-popt}
		\end{subtable}
		
		\bigskip
		\begin{subtable}{\columnwidth}
			\centering
			\caption{Covariances for fit to exp.}
			\begin{tabular}{lrrr}
				\toprule
				&   $p_0$ &   $p_1$ &   $p_2$ \\
				\midrule
  $p_0$ &   7.2e+00 & -2.3e+00 &  1.8e-01  \\
  $p_1$ &  -2.3e+00 &  7.1e-01 & -5.6e-02  \\
  $p_2$ &   1.8e-01 & -5.6e-02 &  4.4e-03  \\
				\bottomrule
			\end{tabular}
			\label{tab:photo-popt2}
		\end{subtable}
		
		\bigskip
		\begin{subtable}{\columnwidth}
			\centering
			\caption{Covariances for fit to sim.}
			\begin{tabular}{lrrr}
				\toprule
				&   $p_0$ &   $p_1$ &   $p_2$ \\
				\midrule
  $p_0$ &   4.3e+00 & -1.4e+00 &  1.0e-01 \\
  $p_1$ &  -1.4e+00 &  4.2e-01 & -3.3e-02 \\
  $p_2$ &   1.0e-01 & -3.3e-02 &  2.5e-03 \\
				\bottomrule
			\end{tabular}
			
			\label{tab:photo-popt3}
		\end{subtable}
	\end{table}
	
	\FloatBarrier
	\section*{References}
	\bibliography{bib}
\end{document}